\documentclass[reprint,aps,prc,twocolumn,longbibliography,superscriptaddress,amsmath,amssymb]{revtex4-2}
\usepackage{epsfig,amsmath,graphicx,amssymb,tabularx}
\usepackage[bookmarks=true,colorlinks,citecolor=blue,linkcolor=blue,anchorcolor=blue,filecolor=blue,urlcolor=blue]{hyperref}
\usepackage[T1]{fontenc}
\usepackage[utf8]{inputenc}
\usepackage{dcolumn}
\usepackage{bm}
\usepackage{multirow}
\makeatletter
\def\NAT@def@citea{\def\@citea{\NAT@separator}}
\makeatother

\begin{document}
\title{Impact of tensor forces on quasifission product yield distributions}
\author{Liang Li}
\affiliation{School of Nuclear Science and Technology, University of Chinese Academy of Sciences, Beijing 100049, China}
\author{Lu Guo}
\email{luguo@ucas.ac.cn}
\affiliation{School of Nuclear Science and Technology, University of Chinese Academy of Sciences, Beijing 100049, China}
\author{K. Godbey}
\email{godbey@frib.msu.edu}
\affiliation{Facility for Rare Isotope Beams, Michigan State University, East Lansing, Michigan 48824, USA}
\author{A. S. Umar}
\email{umar@compsci.cas.vanderbilt.edu}
\affiliation{Department of Physics and Astronomy, Vanderbilt University, Nashville, Tennessee 37235, USA}
\date{\today}
\begin{abstract}
\edef\oldrightskip{\the\rightskip}
\begin{description}
\rightskip\oldrightskip\relax
\setlength{\parskip}{0pt} 
  \item[Background] Quantum shell effects are crucial for the stability and structure of atomic nuclei and play a key role in the discovery of superheavy elements. However, during nuclear collisions, the dynamical evolution of these shell effects disrupts the equilibration process necessary for forming a compound nucleus, leading to the breakup of the initial composite as a result of quasifission. As such, quasifission reactions hinder the formation of a superheavy element (SHE).
  \item[Purpose] In our previous study~[\href{https://doi.org/10.1016/j.physletb.2022.137349}{Phys. Lett. B 833, 137349 (2022)}], we identified that tensor forces significantly influence the evolution of dynamical shell effects in quasifission reactions of the $^{48}\mathrm{Ca}+{}^{249}\mathrm{Bk}$ system, particularly affecting the competition between spherical and deformed shell gaps. This paper aims to extend our previous findings by employing TIJ parametrizations of the Skyrme interaction to investigate the effects of isoscalar and isovector tensor coupling constants on quasifission products. Additionally, we aim to validate our conclusions by analyzing the $^{48}\mathrm{Ti}+{}^{238}\mathrm{U}$ system, for which experimental data is available.
  \item[Method] We employ the microscopic time-dependent Hartree-Fock (TDHF) theory to study the $^{48}\mathrm{Ca}+{}^{249}\mathrm{Bk}$ and $^{48}\mathrm{Ti}+{}^{238}\mathrm{U}$  systems, taking into account the dependence on orientation for deformed
  nuclei and full range of impact parameters. By analyzing fragment distributions
  of neutron and proton numbers, we assess the influence of different isoscalar and isovector tensor coupling
  constants of the effective nucleon-nucleon interaction.
  \item[Results] The quasifission yield distributions of $^{48}\mathrm{Ca}+{}^{249}\mathrm{Bk}$ collision system utilizing SLy5t and T31 parametrizations exhibit more  pronounced spherical shell effects compared to those using SLy5, T44 and T62 sets.   Furthermore, within each parametrization group, the distributions for SLy5t and T31 are closely aligned, as are those for SLy5, T44, and T62. Similarly, the yield distributions for the $^{48}\mathrm{Ti}+{}^{238}\mathrm{U}$ system using SLy5t and T31 also reflect the more pronounced spherical shell effects relative to SLy5 and T62, while the charge distribution shows much better agreement with experimental results for the SLy5t and T62 parametrizations compared to SLy5 and T31.
  \item[Conclusions] The yield distributions for the $^{48}\mathrm{Ca}+{}^{249}\mathrm{Bk}$ and $^{48}\mathrm{Ti}+{}^{238}\mathrm{U}$ systems, when compared across the SLy5, SLy5t, T31, T44, and T62 parametrizations, indicate that the influence of tensor forces on quasifission fragments is reflected in the prominence of shell effects. This influence appears to be sensitive only in specific regions within the isoscalar and isovector coupling constant parameter space. In the $^{48}\mathrm{Ti}+{}^{238}\mathrm{U}$ system, the prominence of shell effects is manifested not only through  shifts in peak positions but also through narrower yield distributions.

\end{description}
\end{abstract}
\maketitle

\section{INTRODUCTION}
In the description of bound fermionic quantum systems, the significance of shells
is paramount as they dictate the system's stability. This is particularly
evident in nuclear physics, wherein protons and neutrons occupy discrete energy levels known as shells.
The gaps between these shells, often referred to as shell closures, lead to the identification of ``magic numbers''. Nuclei with fully occupied shells, termed ``magic nuclei'', exhibit significantly  increased binding energies and  enhanced stability compared to non-magic nuclei~\cite{mayer1948,myers1966}.
The introduction of spin-orbit coupling in the nuclear shell model by Mayer~\cite{mayer1949} and Jensen~\cite{haxel1949} was critical to the correct prediction of known magic numbers $Z=2$, 8, 20, 28, 50, and 82 for protons, and $N=2$, 8, 20, 28, 50, 82, and 126 for neutrons. Next neutron shell closures are predicted potentially at
$N=172$ or $N=184$, while proton closures are expected near $Z=114$ and
potentially at $Z=120$, $Z=124$, or
$Z=126$~\cite{sobiczewski1966,cwiok1996,bender1999,kruppa2000}. These predicted shell closures serve as theoretical cornerstones for the ongoing quest to discover superheavy elements (SHE), particularly in the so-called ``island of stability'', where nuclei with longer half-lives may reside.

The synthesis of SHE remains one of the most compelling areas at the
forefront of nuclear science. The completion of the seventh row of the periodic table, marked by the discovery of element
Oganesson~\cite{oganessian2006} through hot
fusion reactions between $^{48}\mathrm{Ca}$ and ${}^{249}\mathrm{Cf}$, stands as a significant milestone.
However, despite this achievement, experimental efforts to push the limits of the periodic table into the eighth row by synthesizing elements 119 and 120 have yet to succeed~\cite{oganessian2009,hofmann2016,khuyagbaatar2020,tanaka2022}.
One of the major challenges in producing elements with atomic numbers greater than 118 is the scarcity of sufficient trans-Cf isotopes to use as target materials. As a result, researchers are forced to use heavier nuclei than $^{48}\mathrm{Ca}$ for projectile beams. The use of heavier projectiles, such as $^{50}\mathrm{Ti}$ or $^{54}\mathrm{Cr}$, alters the dynamics of the fusion and quasifission process, leading to lower cross-sections and making the synthesis of new elements even more challenging. Additionally, competing reaction channels such as fusion-fission and quasifission~\cite{vardaci2019} pose significant hurdles, necessitating a deeper understanding of their underlying mechanisms. Unlike fusion, which results in the formation of a compound nucleus, quasifission occurs when  two colliding nuclei overcome the capture barrier but fail to form a fully equilibrated compound nucleus, instead re-separating into two  fragments. This non-equilibrium process typically involves shorter interaction times compared to  fusion-fission,  considerable mass transfer, strong mass-angle correlations, and partial retention of the initial properties from the entrance channel~\cite{back1983,shen1987}.

Quasifission reaction dynamics are governed by a complex interplay between entrance and exit channels, including factors such as collision energy~\cite{nishio2012}, deformation and orientation of the colliding nuclei~\cite{hinde1995}, neutron excess in the composite system~\cite{hammerton2015}, and exit channel shell effects~\cite{chizhov2003,wakhle2014,pal2024}. These factors can individually or collectively influence the trajectory and outcome of quasifission. For instance, quasifission events involving actinide targets displayed
a mass-asymmetric yield peak near the $^{208}$Pb mass region at around Coulomb
barrier beam energies~\cite{gippner1986,itkis2004,kozulin2010}. Recent
experiment that measured both the atomic number and mass of quasifission
fragments, have demonstrated the critical influence of proton shell closure at $Z=82$ on
fragment formation~\cite{morjean2017}.

Accurately modeling the quasifission mechanism remains a significant challenge for theoretical approaches due to the intricate interplay between the entrance and exit channels. There are several theoretical approaches describing quasifission through multi-nucleon transfer processes, each capturing different facets of this complex reaction. These frameworks range from phenomenological models such as  the dinuclear system (DNS) model and the multidimensional Langevin equation to microscopic theories like the improved quantum molecular dynamics (ImQMD) and the time-dependent Hartree-Fock (TDHF) theory. The DNS model~\cite{adamian2003,zhu2016,feng2009a,wang2012,guo2017}, treats the system as two interacting nuclei that retain their identities during the reaction, allowing for the exchange of nucleons between them until either fusion or quasifission occurs. This model is effective for analyzing the mass and charge distributions of reaction products. The multidimensional Langevin equation~\cite{zagrebaev2007,schmitt2019,amano2022}, incorporates stochastic terms to account for the energy dissipation and nucleon transfer, providing a framework for studying the influence of friction and fluctuations on the quasifission process. The ImQMD model~\cite{wang2015,zhao2016,li2018} represents nucleons as Gaussian wave packets and introduces nucleon collision in a mean-field nuclear potential, making it suitable for understanding nuclear fragmentation in heavy-ion collisions. Lastly, the TDHF theory~\cite{reinhard2007,guo2007,guo2008,umar2015a,guo2018b,wu2019,wu2020,jiang2020,sun2023,sun2023b} describes the dynamical evolution of nuclear many-body systems with the mean-field approximation, which simplifies the complex many-body wavefunction into a Slater determinant constructed from single-particle wavefunctions. This approach offers a robust framework for simulating and analyzing the dynamics of nuclear reactions near the Coulomb barrier. One advantage of microscopic TDHF calculations is that they require only the parameters of the energy density functional, which are typically fitted to nuclear structure properties, eliminating the need for additional parameters specific to reaction mechanisms. Comparing with the ImQMD model, the TDHF calculations can describe better the structure effects of the nuclear system such as the shell effects and the nuclear shapes in heavy-ion reaction at low incident energies.

Recently, the TDHF theory  has been proven to be a powerful tool for
studying quasifission dynamics~\cite{umar2015c,simenel2018,guo2018,guo2018d,stevenson2019,godbey2020,li2022}. TDHF
calculations exhibit a close agreement with experimental data on mass-angle distributions and final fragment total kinetic energies (TKE)~\cite{kedziora2010,wakhle2014,oberacker2014,goddard2015,hammerton2015,umar2015a,umar2015c,umar2016,prasad2016,wang2016,sekizawa2016,yu2017,li2019,godbey2019,sekizawa2017,sekizawa2019b}.These studies also indicate that spherical shell effects may dominate quasifission
dynamics~\cite{simenel2018,sekizawa2019}.
In addition to spherical shell closures, deformation-induced shell effects, such as those
at $Z=40$ and $Z=52-56$, which have been observed in fission studies, are attributed to octupole deformed shapes in the resulting fission fragments~\cite{scamps2018,scamps2019,bender2020,huang2024}. Despite the distinct differences between fission and quasifission, similar deformation-induced shell effects are also present in quasifission processes~\cite{godbey2019,simenel2021,mcglynn2023,lee2024b}.

The impact of tensor forces on nuclear structure has been extensively studied ~\cite{colo2007,bernard2020,hellemans2012,lesinski2007,sagawa2014}. Notably, research indicates that the tensor force alters the evolution of nuclear shells, having an impact comparable to that of the spin-orbit splitting changes in exotic neutron-rich nuclei~\cite{otsuka2005,otsuka2006,otsuka2010,otsuka2020,bai2009,bai2010}. Despite this, studies examining the contribution of tensor forces in nuclear reactions remain limited due substantial computational costs. Only in recent years have several studies begun to elucidate the influence of tensor forces on fusion barriers and  cross-sections~\cite{guo2018,guo2018b,godbey2019c,sun2022,sun2022c}, as well as on dissipation mechanisms in heavy-ion collisions~\cite{dai2014,stevenson2016}. Our previous study provided the first evidence that tensor forces affect the competition between spherical and deformed shell gaps in quasifission reactions~\cite{li2022}. In this article, we extend our previous study to encompass more Skyrme interaction parametrizations and reaction systems, thereby further investigating the impact of tensor forces in quasifission.

The article is organized as follows. In Sec.~\ref{theory}, we briefly discuss some of the theoretical aspects of our calculations.
Next, in Section~\ref{results} we first present our findings for the $^{48}$Ca+$^{249}$Bk system, which is followed by the
results for the $^{48}\mathrm{Ti}+{}^{238}\mathrm{U}$ system. Finally, a summary and conclusion of the present work is given in Sec.~\ref{summary}.
\begin{figure}[htb!]
	\includegraphics[width=8.6cm]{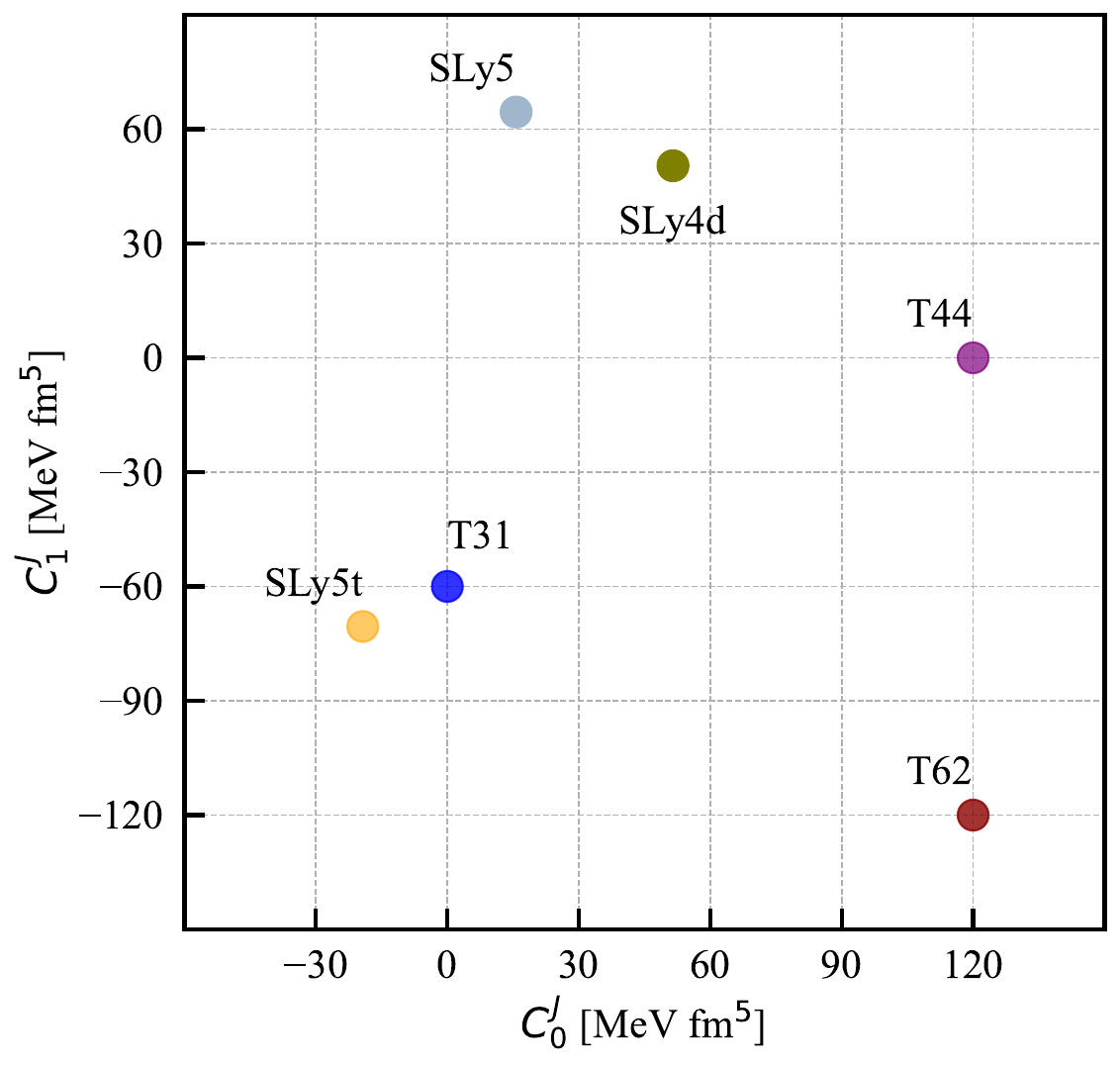}
	\caption{Values of $C^J_0$ and $C^J_1$ for the Skyrme parametrizations explored in this paper.
	}
	\label{fig:C0C1}
\end{figure}
\begin{figure*}[ht!]
    \includegraphics[width=1.5\columnwidth]{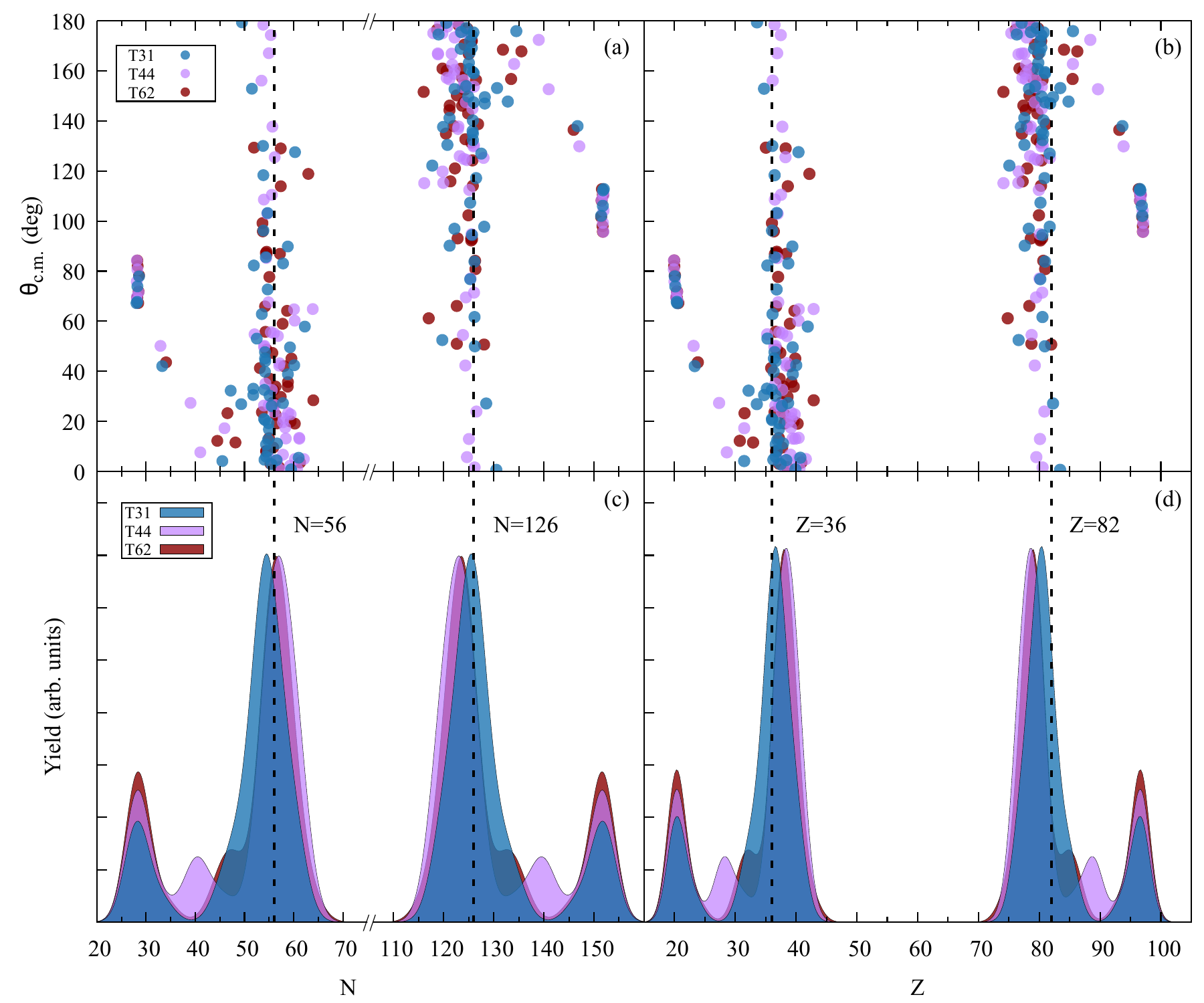}
    \caption{Fragment yields for the $^{48}$Ca+$^{249}$Bk quasifission reaction at the center-of-mass energy $E_{\mathrm{c.m.}}=234$~MeV. The distribution of the scattering angle $\theta_{\mathrm{c.m.}}$  and the fragment yield versus neutron number $N$ (panels (a) and (c)) and proton number $Z$ (panels (b) and (d)) are shown. Three different TIJ forces~\cite{lesinski2007} are used for comparison. The vertical dashed lines correspond to the deformed and spherical shell gaps at $N=56, 126$, and $Z=36, 82$.}
    \label{fig1}
\end{figure*}
\begin{figure*}[htb!]
    \includegraphics[width=1.5\columnwidth]{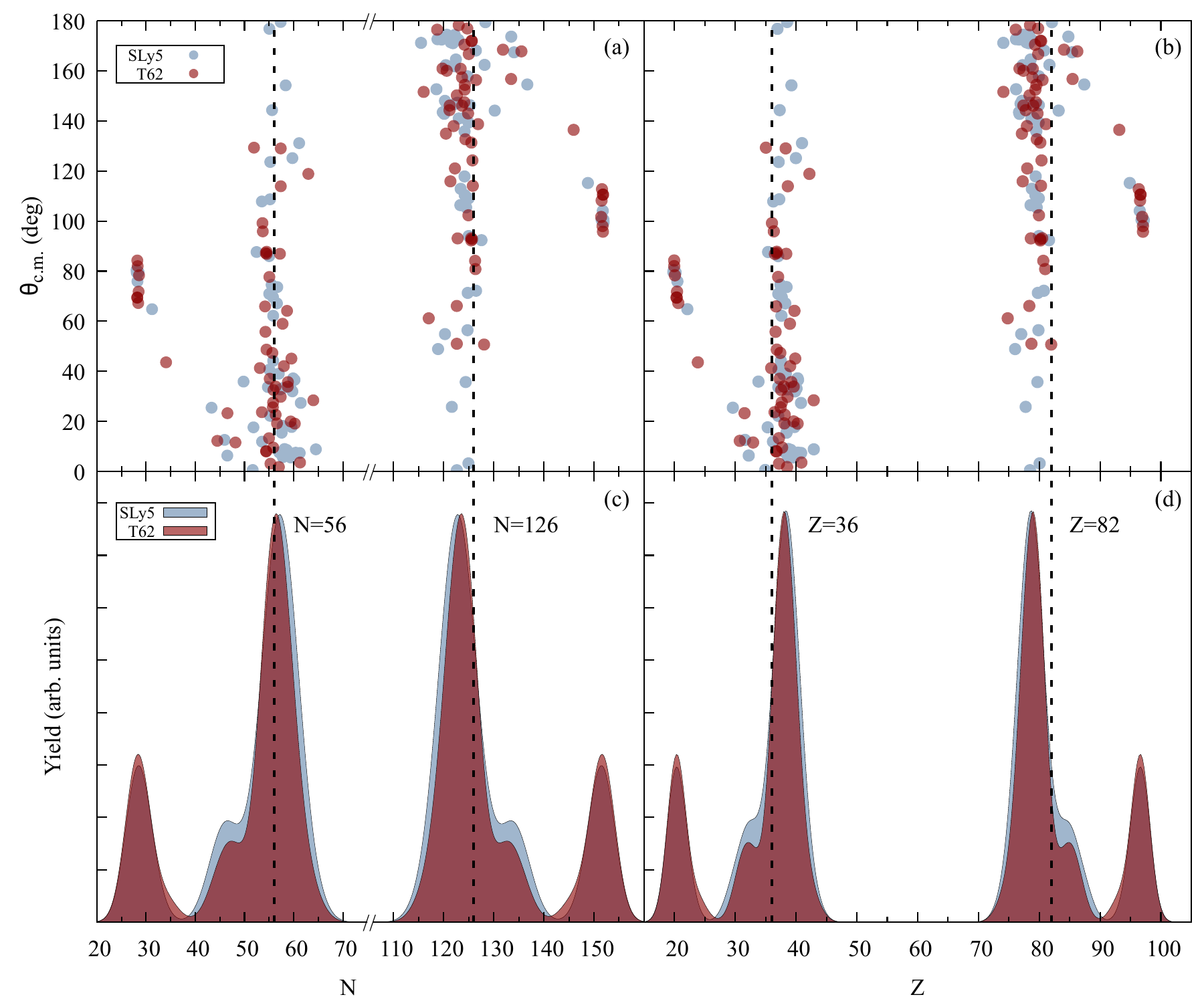}
    \caption{Same as Fig.~\ref{fig1}, except that the slate gray points and shades correspond to results using SLy5 force.
    }
    \label{fig2}
\end{figure*}
\begin{figure*}[ht!]
    \includegraphics[width=1.5\columnwidth]{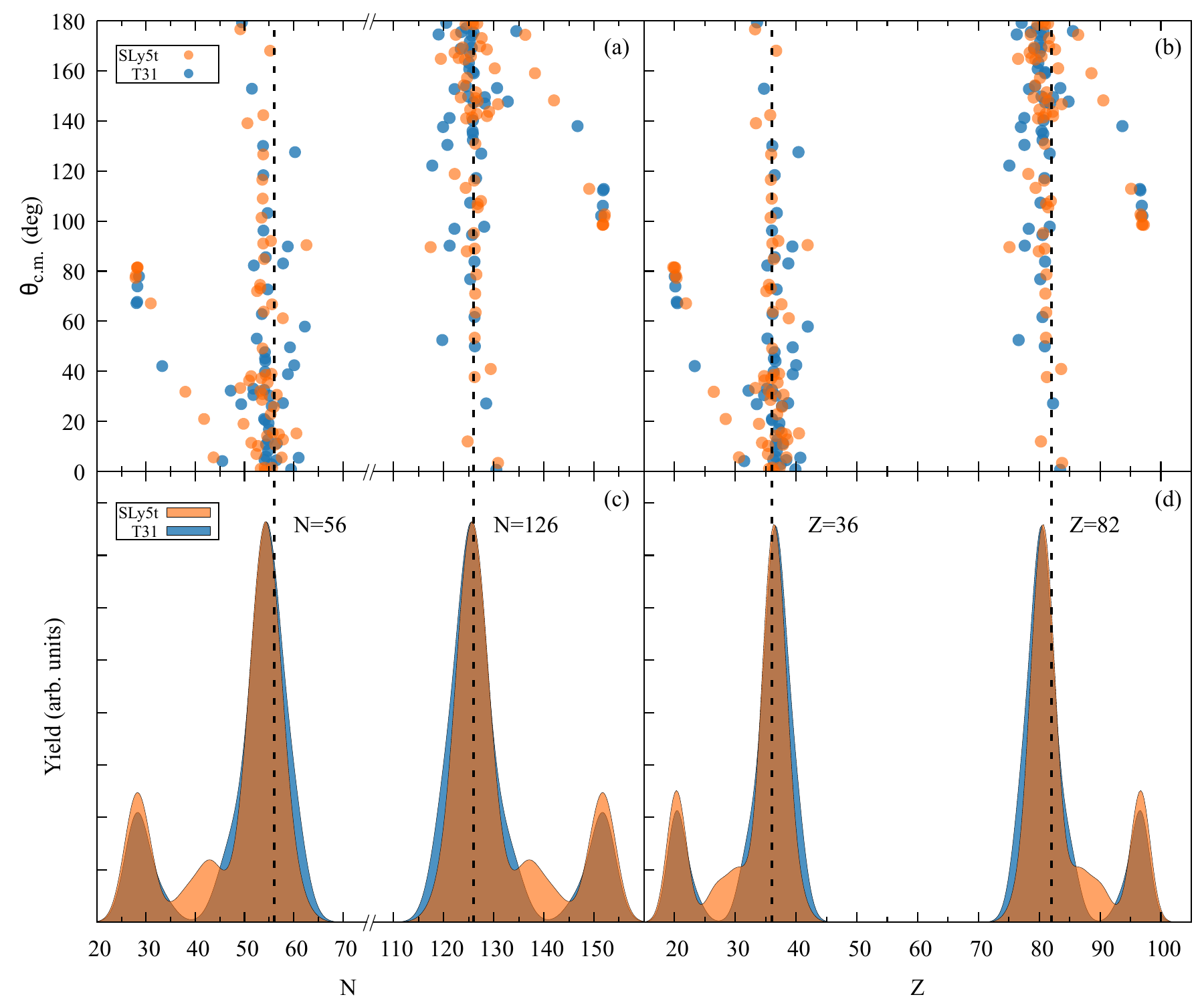}
    \caption{Same as Fig.~\ref{fig1}, except that the orange points and shades correspond to results using SLy5t force.}
    \label{fig:gaussn}
\end{figure*}
\begin{figure}[htb!]
\includegraphics[width=8.6cm]{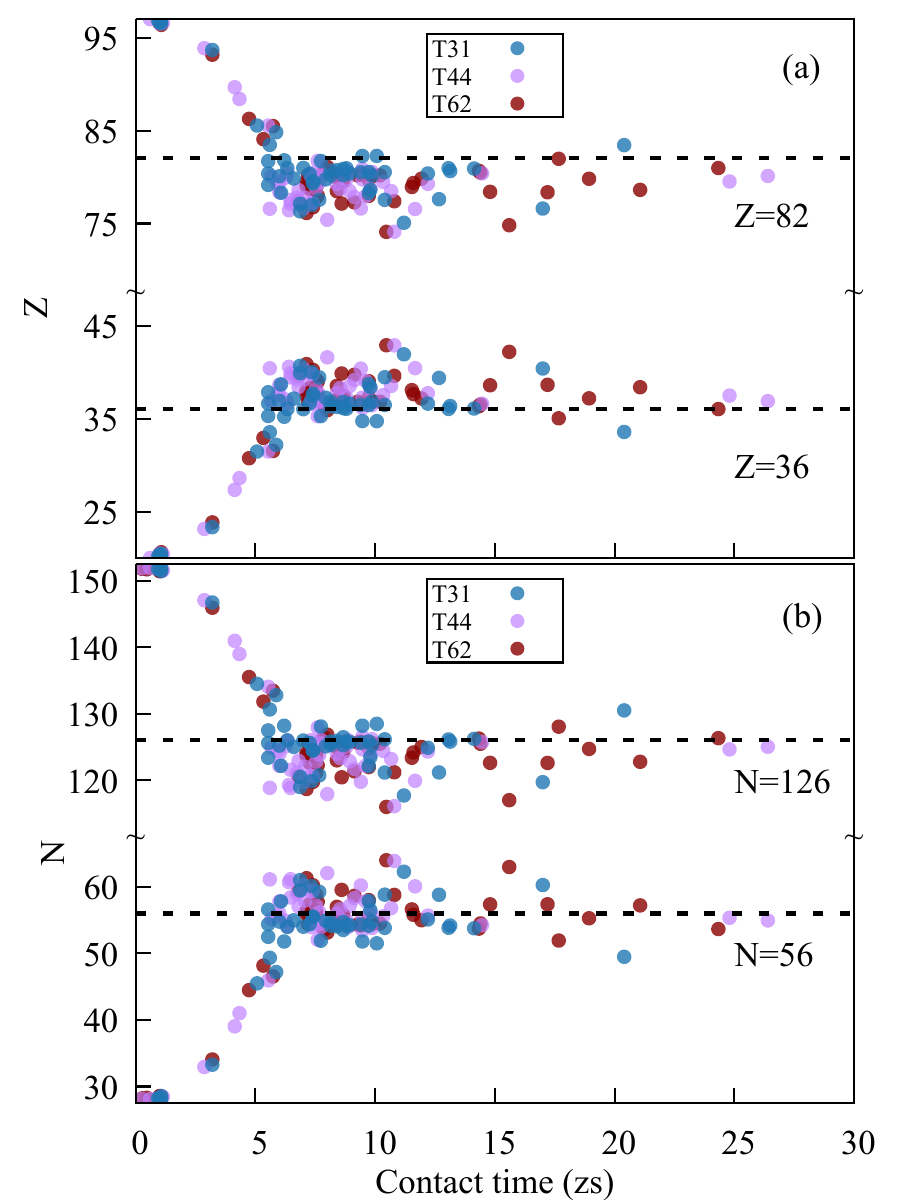}
\caption{(a) Proton number $Z$, (b) neutron number $N$ of quasifission fragments versus contact time using three
TIJ forces~\cite{lesinski2007}. The horizontal dashed lines represent spherical and deformed shell gaps.}
\label{fig:nztime}
\end{figure}
\section{THEORETICAL APPROACH}
\label{theory}
In TDHF applications, the effective nucleon-nucleon interaction utilized is the Skyrme potential~\cite{skyrme1958}, which is used to construct the energy density functional (EDF). The comprehensive Skyrme interaction encompasses central, spin-orbit, and tensor components, with the two-body tensor force expressed as
\begin{align}
v_{\mathrm{T}}&=\dfrac{t_\mathrm{e}}{2}\bigg\{\big[3({\sigma}_\mathrm{1}\cdot\mathbf{k}')({\sigma}_\mathrm{2}\cdot\mathbf{k}')-({\sigma}_\mathrm{1}\cdot{\sigma}_\mathrm{2})\mathbf{k}'^{\mathrm{2}}\big]\delta(\mathbf{r}_\mathrm{1}-\mathbf{r}_\mathrm{2}) \notag \\
&+\delta(\mathbf{r}_\mathrm{1}-\mathbf{r}_\mathrm{2})\big[3({\sigma}_\mathrm{1}\cdot\mathbf{k})({\sigma}_\mathrm{2}\cdot\mathbf{k})-({\sigma}_\mathrm{1}\cdot{\sigma}_\mathrm{2})\mathbf{k}^\mathrm{2}\big]\bigg\} \notag \\
&+t_\mathrm{o}\bigg\{3({\sigma}_\mathrm{1}\cdot\mathbf{k}')\delta(\mathbf{r}_\mathrm{1}-\mathbf{r}_\mathrm{2})({\sigma}_\mathrm{2}\cdot\mathbf{k})\notag \\
&-({\sigma}_\mathrm{1}\cdot{\sigma}_\mathrm{2})\mathbf{k}'
\delta(\mathbf{r}_\mathrm{1}-\mathbf{r}_\mathrm{2})\mathbf{k}\bigg\}.
\end{align}
The constants $t_\mathrm{e}$ and $t_\mathrm{o}$ denote the magnitudes of triplet-even and triplet-odd tensor forces, respectively.

This comprehensive Skyrme interaction is often simplified due to computational constraints. Initial studies neglected the spin-orbit interaction, resulting in anomalies in fusion studies that were later resolved by including spin-orbit terms~\cite{umar1986a,reinhard1988,umar1989}. Until recently, only a portion of dynamical calculations incorporated tensor forces, the majority of which utilized the SLy5t parametrization. The SLy5t interaction~\cite{colo2007} includes the tensor component that was added perturbatively to the previously established SLy5 force~\cite{chabanat1998a}, without fully readjusting the other Skyrme interaction terms.

An alternative approach for constructing a Skyrme  interaction parametrization that includes tensor forces involves a complete refitting of all interaction parameters. This technique was employed in Ref.~\cite{lesinski2007}, resulting in the development of the TIJ family of Skyrme forces, which comprises 36 distinct parametrizations. The TIJ forces adhered to a fitting protocol similar to that of the SLyN forces, with the incorporation of the tensor component. Both methods of including tensor forces have strengths and limitations. A Skyrme parametrization without fully readjusting the other Skyrme interaction terms, like SLy5t, only offers an optimal parametrization for a limited set of observables, but it is easier to isolate and compare the influence of the tensor force in dynamical reactions. TIJ parametrizations provide a broad coverage of the isoscalar tensor coupling constant $C_0^J$ and isovector tensor coupling constant $C_1^J$. However, the extensive number of parametrizations makes the TIJ family less practical for use in dynamical calculations as the reaction calculations are computationally demanding. The relationship between the coupling constants  $C_0^J$ and $C_1^J$ and the Skyrme interaction parameters is given by
\begin{equation}
	\begin{aligned}
   C_0^J &= \frac{1}{16}(t_1-t_2) -\frac{1}{8}(t_1 x_1+t_2 x_2)+ \frac{5}{16}t_e+\frac{15}{16}t_o,\\
    C_1^J &= \frac{1}{16}(t_1-t_2) - \frac{5}{16}t_e+\frac{5}{16}t_o.	\label{eq:C0C1}
\end{aligned}
\end{equation}
 For a more detailed understanding of the SLy5t and TIJ parameters, please refer to Ref.~\cite{colo2007} and ~\cite{lesinski2007}.

This paper presents the evidence of the tensor force's impact on shell effects in quasifission reactions using the TDHF theory. The TDHF equations, which describe the time-dependent behavior of the system, are derived through variation of the many-body wavefunction, $\Phi(t)$, constructed from single-particle wavefunctions $\phi_{\lambda}$,
\begin{equation}
h(\{\phi_{\mu}\})\ \phi_{\lambda} (r,t) = i \hbar \frac{\partial}{\partial t} \phi_{\lambda} (r,t)
            \ \ \ \ (\lambda = 1,...,A) \ ,
\label{eq:TDHF}
\end{equation}
where $h$ represents the resulting single-particle Hamiltonian derived from the effective interaction. In the last decade, it has become computationally feasible to carry out TDHF calculations on a 3D Cartesian grid without symmetry constraints, using significantly more precise numerical methods~\cite{umar2006c,sekizawa2013,maruhn2014,schuetrumpf2018,abhishek2024}. For this work, TDHF calculations are conducted by positioning the two static HF solutions for the target and projectile 26~fm apart, presuming that the two nuclei reached this distance along a Coulomb trajectory, and a boost is applied to accelerate the nuclei corresponding to a given center of mass energy. The TDHF equations are then used to evolve the nuclei over time as the reaction progresses. In the scenario of quasifission, the time development halts  as soon as the distance between the final  fragments exceeds 26~fm. The final scattering angle is then determined by extending the evolution to infinity along a Coulomb trajectory. For the dynamic evolution, a numerical box of $60\times28\times48$~fm$^{3}$ is used. The time step is set at 0.2~fm/c and the grid spacing at 1.0~fm.

To position the prolate-deformed $^{249}$Bk and $^{238}$U  nuclei, after calculating the ground state wave functions, a rotation operator is applied to the ground state using the B-spline interpolation method~\cite{pigg2014} to obtain wave functions in various orientations relative to the collision axis. We have examined the orientations of $^{249}$Bk and $^{238}$U at angles of $\beta=0^\circ$, $30^\circ, 45^\circ, 60^\circ, 90^\circ$ within the reaction plane. For each orientation, calculations begin at an impact parameter $b = 0$~fm with an increment $\Delta b = 0.5$~fm until quasielastic reactions occur. Different impact parameters and orientations result in distinct yield contributions. These contributions are quantified using
\begin{equation}
	\sigma_{\lambda} \propto \int_{b_{\text{min}}}^{b_{\text{max}}} b \, \text{d} b \int_{0}^{\frac{\pi}{2}} \text{d} \beta \sin (\beta) P_{b}^{(\lambda)}(\beta)\;,
\end{equation}
where \( \lambda \) denotes a specific reaction channel, and \( P_{b}^{(\lambda)}(\beta) \) is the probability corresponding to a given impact parameter \( b \) and orientation angle \( \beta \). This probability is either 0 or 1 for the reaction channel \( \lambda \).
\section{RESULTS and Discussion}
\label{results}
In this work, we study the quasifission reactions of $^{48}\mathrm{Ca}+{}^{249}\mathrm{Bk}$ system using the TIJ parameter set and compare these results with those obtained using the SLy5 and SLy5t parametrizations. Building on this foundation, we further examine the results for the $^{48}\mathrm{Ti}+{}^{238}\mathrm{U}$ system using the SLy5, SLy5t, and TIJ parameter sets, comparing them with experimental data.
\subsection{$^{48}$Ca+${}^{249}$Bk Results}
Initially, we conducted comprehensive quasifission calculations for the $^{48}$Ca+${}^{249}$Bk system utilizing the SLy5 and SLy5t interactions~\cite{li2022}. The SLy5 interaction replicated the findings of prior calculations using the SLy4d force~\cite{godbey2019}. However, the addition of the tensor force in the SLy5t interaction displayed a marked propensity to favor spherical shell effects over deformed ones. Due to limitations regarding the SLy5t parametrization, specifically that the tensor force parameters were adjusted without recalibrating other components of the Skyrme interaction, we also executed calculations employing the TIJ parametrizations of the Skyrme interaction. Owing to computational time constraints, we selected  three sets of parameters: T31, T44, and T62 from the TIJ family. Figure~\ref{fig:C0C1} displays the values of the coupling constants $C^J_0$ and $C^J_1$ for the Skyrme parametrizations utilized in this paper and previous studies.
From Fig.~\ref{fig:C0C1}, it is evident that the T31 parametrization exhibits an isoscalar coupling constant \(C_0^J\) of 0 MeV fm$^5$ and an isovector coupling constant \(C_1^J\) of -60 MeV fm$^5$. This parametrization is among the closest to the SLy5t interaction in terms of \(C_0^J\) and \(C_1^J\) values among the 36 parameter sets~\cite{lesinski2007}. In contrast, the T62 parametrization employs identical but opposite coupling constant values, with \(C_0^J\) and \(C_1^J\) equal to 120, -120 MeV fm$^5$, respectively. The T44 parametrization, on the other hand, utilizes a positive isoscalar coupling constant \(C_0^J\) of 120 MeV fm$^5$, while the isovector coupling constant \(C_1^J\) is 0 MeV fm$^5$. The variations in \(C_0^J\) and \(C_1^J\) across these parametrizations provide a reference for analyzing the influence of tensor isoscalar and isovector coupling constants on quasifission processes.

Figure~\ref{fig1} displays  the distribution of scattering angle $\theta_{\mathrm{c.m.}}$ versus (a) neutron number $N$, (b) proton number $Z$, as well as fragment neutron number yield (c) and proton number yield (d), for three Skyrme parametrizations with different tensor coupling constants: T31, T44, and T62. Each point in Fig.~\ref{fig1} (a) and (b) illustrates a TDHF outcome for the $^{48}$Ca+${}^{249}$Bk reaction at a specific impact parameter $b$ and orientation $\beta$. Each computational result requires several days of processing using 16 cores on an HPC cluster. As we observe in Fig.~\ref{fig1} (a) and (b), with the exception of the quasielastic arms, the points are highly focused around vertical lines. This is a striking depiction of the shell effects. To more clearly distinguish the differences among these three sets of TIJ results,  we plot the smooth yield distributions of neutron and proton numbers, obtained from various orientations and impact parameters, in Fig.~\ref{fig1} (c) and (d), respectively.
In Fig.~\ref{fig1} (c), the neutron number yield for the T31, T44, and T62 parametrizations is plotted using blue, purple, and dark red shades, respectively. The peak of the blue shade, corresponding to the T31 parametrization, is centered at the vertical dashed line at $N = 126$, which represents the spherical shell number. In contrast, the dark red shaded peak for T62 is closest to the vertical line at $N = 56$, which corresponds to a deformed shell. The purple shaded peak for T44 is very near the dark red T62 peak, although it is slightly further from the deformed shell at $N = 56$. Figure~\ref{fig1} (d)  depicts the fragment yields for proton number, where the peaks for T44 and T62 are also closely aligned, while there is a two proton number difference between these peaks and that of T31. Although the proton peak of heavy fragments for T31 is closer to the spherical shell at $Z=82$ than those for T44 and T62, it is not as pronounced as the neutron peak for T31 at $N=126$. For light fragment proton numbers, the peak for T31 is concentrated near $Z=36$. Recent TDHF calculations have also observed light fragment shell effects in the Kr$-$Zr region~\cite{godbey2019}, while recent fission studies indicate the existence of a deformed proton shell at $Z \approx 36$~\cite{kozulin2022}. Our results further demonstrate that similar deformation-induced shell effects, previously observed in fission, are also present in quasifission.

The shift of the peak from T62 and T44 to T31 in Fig.~\ref{fig1} closely resembles the shift observed from SLy5 to SLy5t in our previous study~\cite{li2022}, suggesting a comparative analysis between the TIJ parametrizations and the SLy5 and SLy5t interactions. Figure~\ref{fig2} presents a comparison between the SLy5 and T62 parametrizations. Although SLy5 was fitted without considering tensor forces and T62 incorporated tensor forces, and their positions are farthest apart in Fig.~\ref{fig:C0C1}, the yield distributions obtained from both are nearly identical. Furthermore, as illustrated in Fig.~\ref{fig1} that the peaks for T62 and T44 are very close, and considering the similar results for SLy4d reported in~\cite{godbey2019}, we observe that the peaks of the yield distributions remain almost unchanged despite significant variations in the
$C_0^J$ and $C_1^J$ coupling coefficients among SLy4d, SLy5, T44 and T62. In Fig.~\ref{fig:gaussn} we compare the distributions between the SLy5t and T31 parametrizations, where the peaks for SLy5t and T31 are almost perfectly aligned. Although SLy5t and T62 incorporate tensor terms in entirely different ways---SLy5t perturbatively and T62 through a full refit---their $C_0$ and $C_1$ coupling coefficients are remarkably similar. Thus, we can categorize the currently involved Skyrme parameters into two groups: one group consisting of SLy4d, SLy5, T44, and T62, and the other group comprising SLy5t and T31. Results within each group are very similar, while there are significant distribution differences between the two groups. This suggests that the effects of tensor forces on the prominence of shell gaps in quasifission may be sensitive only in specific regions of the $C_0^J$ -- $C_1^J$ coupling coefficient map, while remaining insensitive in others.

An alternative perspective on shell effects involves plotting neutron and proton distributions as a function of contact time during the quasifission process. This method allows for the examination of the dependence of shell gaps on the reaction's time scale.
Figure~\ref{fig:nztime} shows the neutron and proton numbers of the heavy and light fragments as a
function of the contact time for the $^{48}\mathrm{Ca}+{}^{249}\mathrm{Bk}$ reaction.
The contact time is defined as the time elapsed from the first touching
of the target and projectile to the separation of the fragments (density overlap of about
$0.03$~fm$^{-3}$).
In Fig.~\ref{fig:nztime} (a), we observe that the proton number of the heavy fragments predominantly fall below the $Z=82$ line. The blue dots representing the T31 parametrization are systematically closer to this line, which consequently affects the proton number of the light fragments, concentrated in the shell gap of the $Z=36$ (krypton) region.
In Fig.~\ref{fig:nztime} (b), the
neutron numbers of the heavy fragments shift upward towards the $N=126$ shell closure for the T31, in contrast to those for the T44 and T62 parametrizations. This shell effect is reversed for the $N=56$ deformed shell
gap, where most T31 points fall bellow the $N=56$ line. Another observation from Figure~\ref{fig:nztime} is that significant fragments near the shell gaps outlined by the horizontal dashed lines appear only when the contact time exceeds 5 zs. This time scale is consistent with the quasifission timescales observed in experiments~\cite{shen1987}. One conclusion we can draw from these observations is that, despite variations in impact parameters and deformation orientations, generally the equilibration process is interrupted by shell effects when contact times exceed 5 zs. With a fixed number of neutrons and protons in the system, competing shell gaps influence each other in the formation of heavy and light fragments. From these we may conclude that spherical
shell gaps compete with the deformed ones for fragment formation and that this competition is influenced by the tensor component underlying effective interaction.

\begin{figure}[htb!]
	\includegraphics[width=8.6cm]{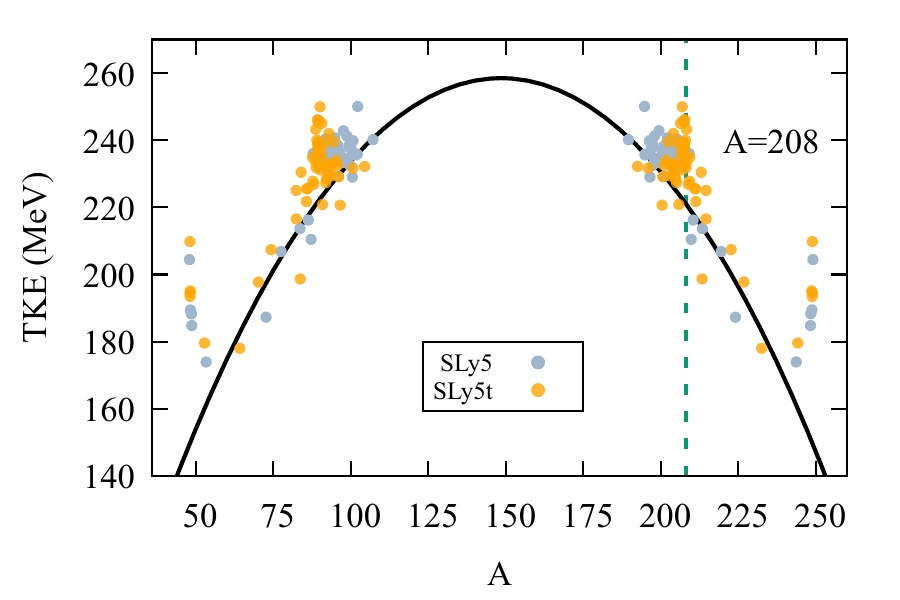}
	\caption{TKE-mass distribution of QF fragments in $^{48}\mathrm{Ca}+{}^{249}\mathrm{Bk}$ using SLy5 and SLy5t forces. The solid line is the Viola systematics.}
	\label{fig:CaBktke}
\end{figure}
In experimental studies of nuclear reactions, the correlations between mass and total kinetic energy
(TKE) of fragments are often used to distinguish the quasielastic events from fully damped
events such as quasifission and fusion-fission. In TDHF, the kinetic energy of well-separated fragments can
be simply calculated from the knowledge of the fragments' mass and velocity. The asymptotic TKE value is then obtained by summing the kinetic energy of the fragments with their Coulomb
potential energy. Figure~\ref{fig:CaBktke} displays the
TKE-mass distribution of quasifission fragments of $^{48}\mathrm{Ca}+{}^{249}\mathrm{Bk}$ collisions employing the SLy5 (slate gray
dots) and the SLy5t (orange dots) Skyrme forces. The Viola systematics (solid line) is included for reference. The TKE are distributed around the Viola estimates, indicating that most of the relative kinetic energy has been dissipated in the collision. The TKE distributions for the T31, T44 and T62 parametrizations are similar to those of SLy5 and the SLy5t in Fig.~\ref{fig:CaBktke}, and therefore, for simplicity, they are not shown in the figure.

The investigation of shell effects at the single-particle level in dynamical quasifission reactions presents significant challenges. The complexity arises due to the range of data points involved, which result from variations in impact parameters and orientation angles. This situation is akin to mass or charge distributions observed in experimental quasifission or asymmetric fission studies, where shell effects are inferred from peaks in these distributions. In contrast, static nuclear structure studies provide a single set of results for a given Skyrme EDF, or in traditional fission barrier calculations as a function of deformation parameters, where gap evolution can be clearly observed from a single computation. In dynamical processes, the situation is further complicated by the inclusion of the system's excitation energy in the single-particle levels, making it difficult to determine their true energies. In our previous work~\cite{li2022}, we addressed this by removing the excitation energy from single-particle levels. This involved selecting a point near the SLy5 peak, analyzing post-scission fragments, and calculating static internal energies under quadrupole and octupole deformation constraints, then repeated the calculations with the SLy5t force. We found that the tensor force enhances the neutron $N=126$ gap in the heavy fragment, while the neutron gaps in the light fragment remain mostly unchanged, except for a shift. For protons, there was a slight enhancement of the $Z=82$ gap in the heavy fragment and a more pronounced enhancement of the $Z=36$ gap in the light fragment. We recognized this treatment as a very rough approximation of the actual situation, referring to it as "a limited study"

\begin{figure}[ht!]
	\includegraphics[width=8.6cm]{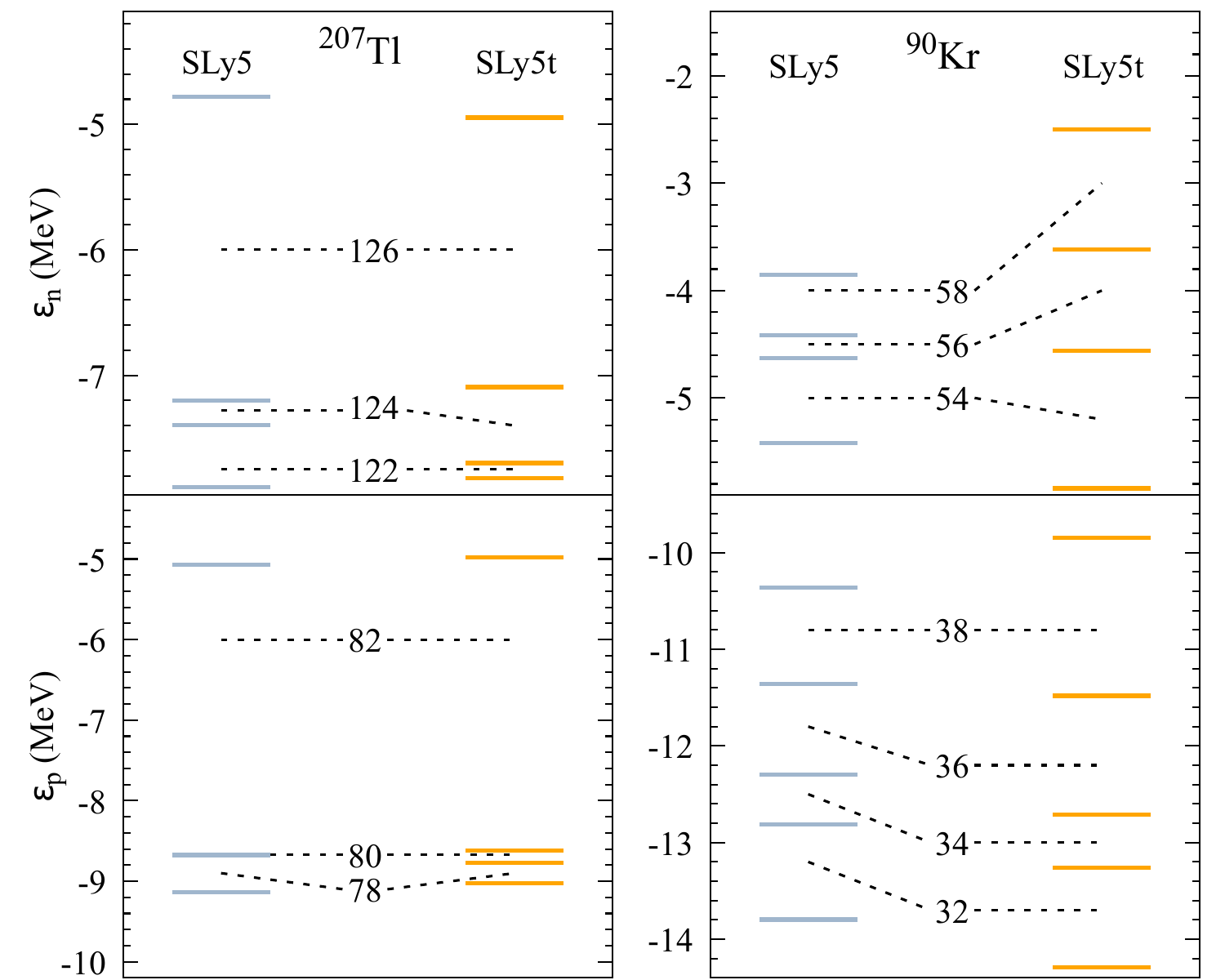}
	\caption{Single-particle energy levels for the heavy (${ }^{207} \mathrm{Tl}$) and light (${ }^{90} \mathrm{Kr}$) fragments with SLy5 and SLy5t forces for neutrons (upper panel) and for protons (lower panel).}
	\label{fig:spe_TlKr}
\end{figure}

In this work, we adopt an opposite approach: we select a point near the SLy5t peak, constrain quadrupole and octupole deformations, and perform the same calculations using SLy5. Figure~\ref{fig:spe_TlKr} presents the changes of single-particle levels resulting from this treatment. As anticipated, the $Z=82$ and $Z=36$ proton gaps, along with the $N=54$ neutron gap using SLy5 are suppressed compared to those from SLy5t. However, the neutron $N=126$ gap unexpectedly exhibits a slight increase. When selecting another data point from the available options, the single-particle energy levels may vary, resulting in outcomes that could partially align with our expectations or deviate from them. Accurately describing single-particle levels with a Skyrme EDF is challenging even in nuclear structure studies, and this difficulty is amplified in dynamical nuclear reactions, necessitating further simplifications and approximations. Thus, the observed enhancements or suppressions of shell gaps in single-particle level diagrams should be regarded as supplementary evidence for our previous conclusions about the influence of tensor forces on shell effects in quasifission, rather than as definitive indicators of their impact.
\begin{figure*}[ht!]
	\includegraphics[width=1.8\columnwidth]{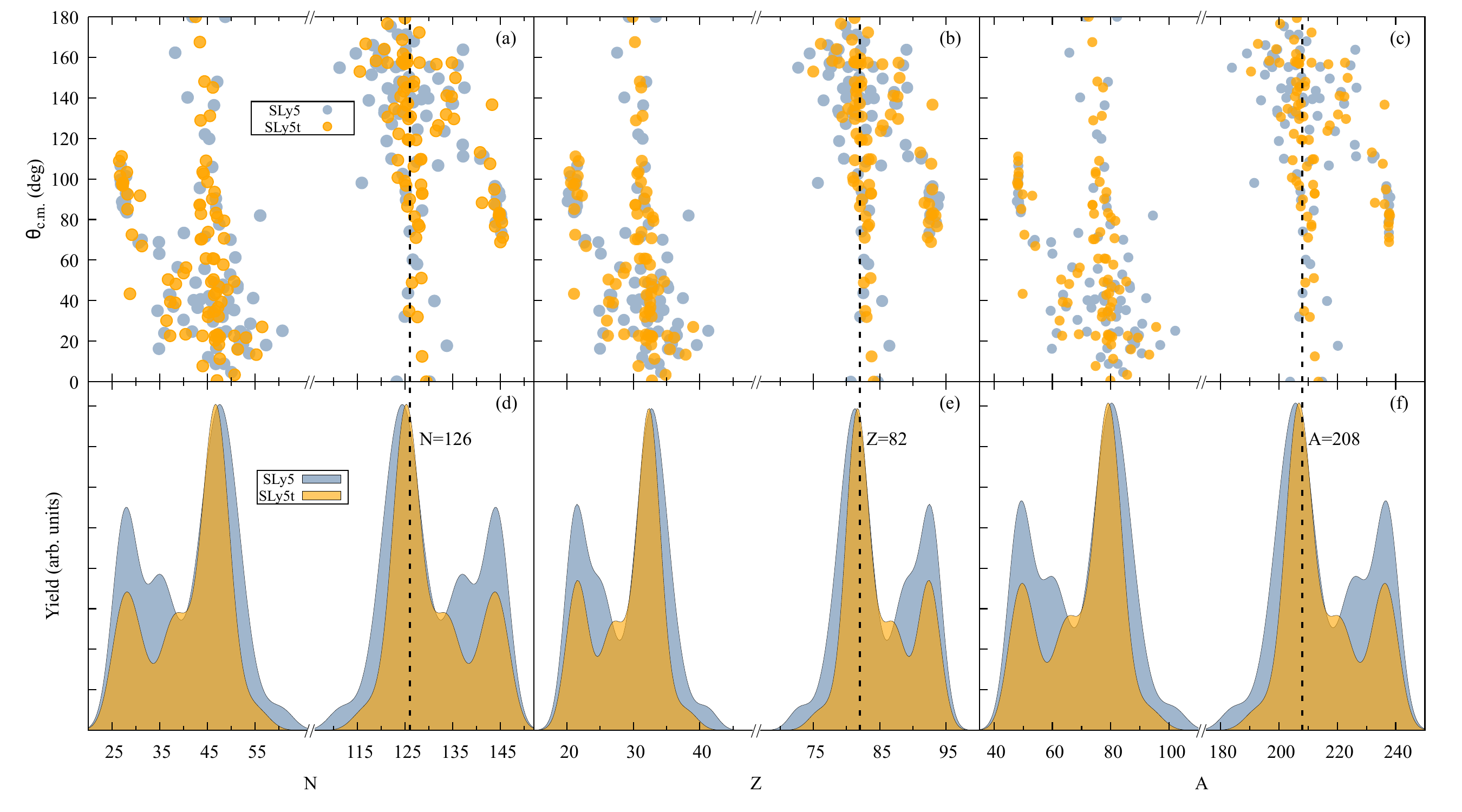}
	\caption{Quasifission reaction for $^{48}\mathrm{Ti}+{}^{238}\mathrm{U}$ at the center-of-mass energy $E_{\mathrm{c.m.}}=229.68$~MeV.  The distribution of the scattering angle $\theta_{\mathrm{c.m.}}$  and the fragment yield versus neutron number $N$  (panels (a) and (d)), proton number $Z$  (panels (b) and (e)), and mass number $A$  (panels (c) and (f)), respectively. The SLy5t force (orange
		shade) is compared with SLy5 force (slate gray shade). The vertical dashed lines correspond to the spherical shell $N=126$,$Z=82$ and mass number $A=208$.}
	\label{fig:gaussn_TiU_1}
\end{figure*}
\begin{figure}[ht!]
	\includegraphics[width=8.6cm]{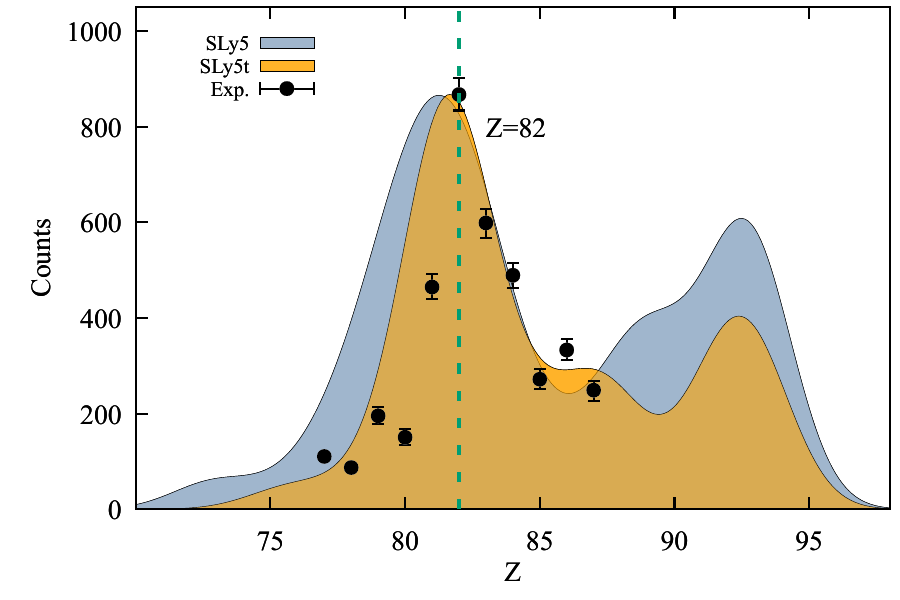}
	\caption{Charge distribution for the quasifission reactions of $^{48}\mathrm{Ti}+{}^{238}\mathrm{U}$ using the
		SLy5 and SLy5t forces. The experimental data is from~\cite{morjean2017}.}
	\label{fig:morjeansly}
\end{figure}
\begin{figure*}[ht!]
	\includegraphics[width=1.8\columnwidth]{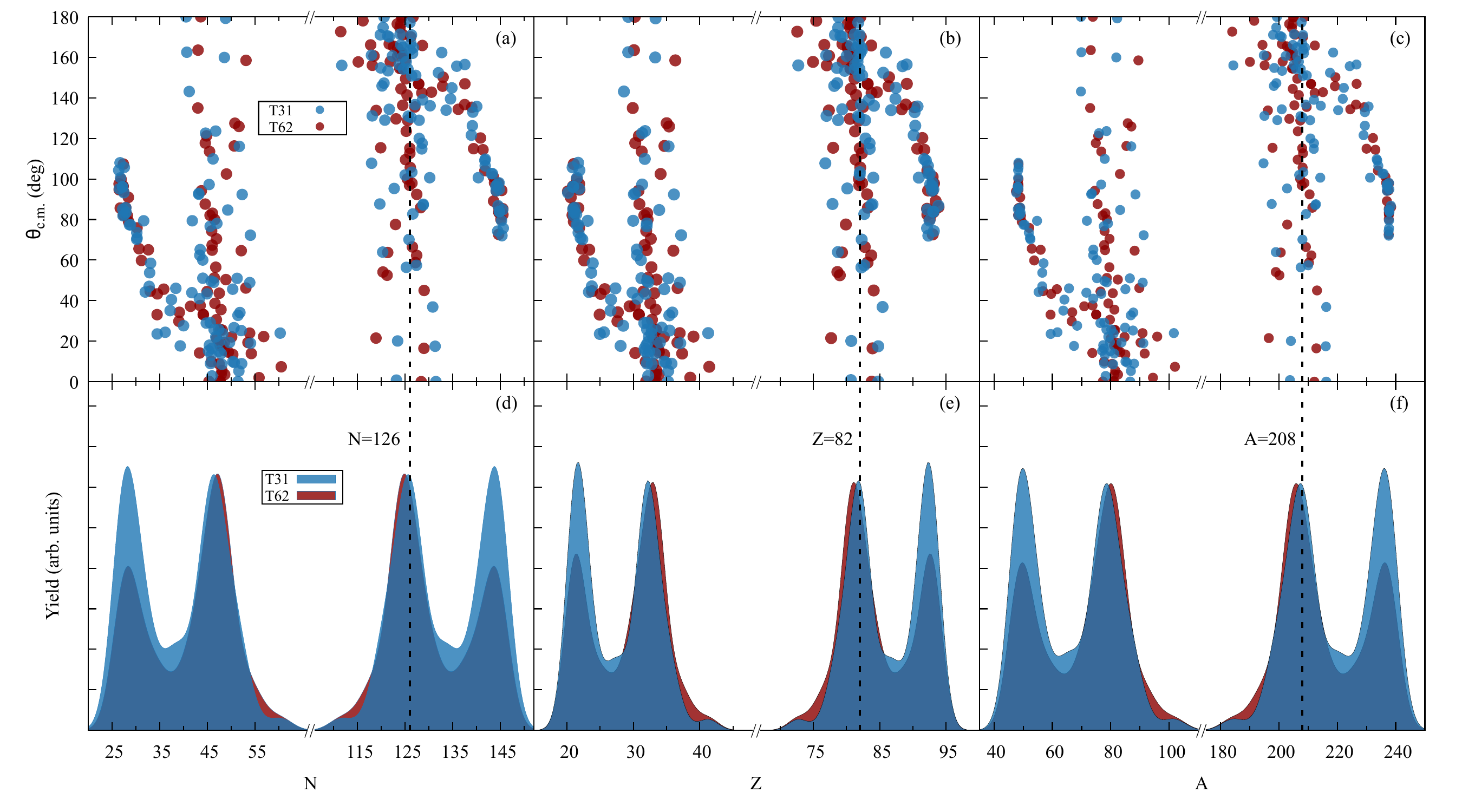}
	\caption{Same as Fig.~\ref{fig:gaussn_TiU_1}, except that the comparison is between T31  (blue
		shade) and T62  (dark-red
		shade) forces.}
	\label{fig:gaussn_TiU_Tij}
\end{figure*}
\subsection{$^{48}$Ti+${}^{238}$U Results}

\begin{figure}[ht!]
	\includegraphics[width=8.6cm]{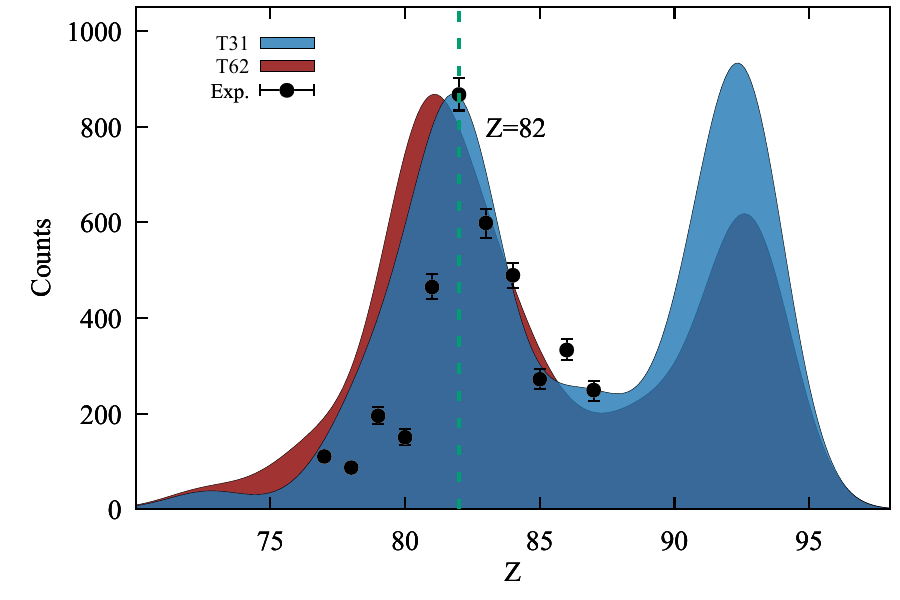}
	\caption{Same as Fig.~\ref{fig:morjeansly}, except that the comparison is between T31 and T62 forces}
	\label{fig:morjeanT62}
\end{figure}

To validate our conclusions across different systems and to enable comparison with experimental data, we also performed calculations for the $^{48}\mathrm{Ti}+{}^{238}\mathrm{U}$ system studied
experimentally in Ref.~\cite{morjean2017} at the center-of-mass energy $E_{\mathrm{c.m.}}=229.68$~MeV. Due to the prolate shape of $^{48}$Ti nucleus, as opposed to the spherical nature of $^{48}$Ca nucleus, the reaction orientations are more complex. To account for the orientation effects more accurately, we consider two specific orientations of $^{48}$Ti nucleus: one with their long axis aligned parallel to the collision direction and another perpendicular to it. This approach doubles the computational effort for the  $^{48}\mathrm{Ti}+{}^{238}\mathrm{U}$ system compared to the  $^{48}\mathrm{Ca}+{}^{238}\mathrm{Bk}$ collision. To conserve computational resources, we omit the T44 force used in the $^{48}\mathrm{Ca}+{}^{238}\mathrm{Bk}$ calculations, as previous results indicated minimal differences between T44 and T62. Therefore, we focus on computing the four Skyrme parameter sets: SLy5, SLy5t, T62, and T31.

In Fig.~\ref{fig:gaussn_TiU_1} we compare the
distributions of scattering angle $\theta_{\mathrm{c.m.}}$ and yield with and without the tensor force for the SLy5 and Sly5t interactions. The doubling of orientations leads to a broader distribution of data points, making the distinction between quasielastic and quasifission events less clear compared to the $^{48}\mathrm{Ca}+{}^{238}\mathrm{Bk}$ system. The data points for elastic and quasielastic events calculated using different Skyrme EDFs may vary in number, leading to significant differences in quasielastic yield heights for SLy5 and SLy5t in Fig.~\ref{fig:gaussn_TiU_1}. However, since our focus is on the peaks associated with quasifission, this discrepancy does not affect our discussion. In Fig.~\ref{fig:gaussn_TiU_1} (d) and (e), the peak centers for the SLy5t parameter set, which includes tensor force, are positioned closer to the spherical shell closures at $N=126$ and $Z=82$ compared to the SLy5  set without tensor force. Although the observed shift is smaller, the movement of the peak relative to the shell positions is consistent with the behavior observed for SLy5 and SLy5t in the $^{48}\mathrm{Ca}+{}^{238}\mathrm{Bk}$ collision~\cite{li2022}, indicating a stronger manifestation of spherical shell effects when the tensor force is included.

In addition to the shift in peak position, a notable difference in Fig.~\ref{fig:gaussn_TiU_1} (d), (e) and (f) is the width of the quasifission yields.
A key insight is that a more vertically concentrated arrangement of data points results in a narrower distribution, whereas  a broader horizontal spread produces a wider distribution. The quasifission yields for SLy5t, centered around the vertical dashed lines at $N=126$ and $Z=82$, are narrower than those for SLy5, suggesting a greater production of fragments near these spherical shells in the SLy5t interaction.
Thus, the width of the fragment yield distribution, centered around the vertical lines representing shell closures, is directly related to the strength of shell effects; a narrower distribution signifies a stronger shell effect. This observation further illustrates the role of tensor forces in shell effects, distinct from their influence on the previously noted shifts in peak centroids. It further indicates that in the $^{48}\mathrm{Ti}+{}^{238}\mathrm{U}$  system, the tensor force enhances the prominence of spherical shell effects just from a distinct perspective. For a more detailed comparison, including the experimental data from~\cite{morjean2017}, we plot the charge distribution for SLy5 and SLy5t in Fig.~\ref{fig:morjeansly}. In this figure, we aligned quasifission yield peaks of SLy5 and SLy5t and experimental maximum values horizontally. This expanded display reveals that the yield for the SLy5t interaction, incorporating tensor force, aligns more closely with experimental data than that for SLy5 interaction.

In Fig.~\ref{fig:gaussn_TiU_Tij}, we illustrate the
distributions of scattering angle and fragment yields for the two tensor interaction (TIJ) forces, T31 and T62. This figure shows a greater number of quasielastic points, resulting in quasielastic peaks that are nearly as pronounced as the quasifission peaks. In contrast to Fig.~\ref{fig:gaussn_TiU_1}, where SLy5 and SLy5t display a clear difference in the width of the quasifission yield distribution, the quasifission yield distributions for T31 and T62 are nearly identical in width, with only differences in the peak positions. The peak of the T31 parametrization aligns closely with the spherical shell closures at $N=126$ and $Z=82$, while the peaks for T62 exhibit a leftward shift from these values. This indicates that the spherical shell effects are more pronounced in quasifission fragments of $^{48}\mathrm{Ti}+{}^{238}\mathrm{U}$ system when utilizing T31 compared to T62. Additionally, we present an expanded charge distribution plot for the T31 and T62 interactions alongside the experimental data in Fig.~\ref{fig:morjeanT62}. It is evident that the peak for T31 is centered around Z=82, whereas the peak for T62 is shifted leftward relative to the T31 peak, exhibiting an offset analogous to that observed between SLy5 and SLy5t in Fig.~\ref{fig:morjeansly}. Furthermore, the yield of T31 parametrization matches experimental data better than that for T62. The consistently stronger dominance of spherical shell effects observed in the two groups of Skyrme EDFs---SLy5t and T31 versus SLy5 and T62---across both the $^{48}\mathrm{Ca}+{}^{249}\mathrm{Bk}$ and the $^{48}\mathrm{Ti}+{}^{238}\mathrm{U}$ systems reinforces our previous assertion that the influence of tensor forces on the prominence of shell gaps in quasifission is sensitive only in specific regions of the $C_0^J$ -- $C_1^J$ coupling coefficient map.
\section{SUMMARY AND CONCLUSIONS}
\label{summary}
Quasifission reactions are one of the reactions that occur in the doorway to
superheavy element formation, in addition to being important in their own right.
They occupy the reaction chart between deep-inelastic collisions and
fusion-fission, with time scales many times that of the traditional heavy-ion
collision times. In this sense they are a bridge between collisions that have
little equilibration and full equilibration. Consequently, they allow for the
development of dynamical shell effects that require a longer time scale. The
TDHF theory using EDFs have been very successful in describing the observed
features of these reactions. We are now at a stage where the details of the
EDFs, that are primarily fitted to nuclear structure properties, are being
considered. In this work, TDHF theory was utilized to study quasifission for the
reactions of  $^{48}\mathrm{Ca}+{}^{249}\mathrm{Bk}$ and $^{48}\mathrm{Ti}+{}^{238}\mathrm{U}$,
using the Skyrme EDFs with and without the tensor interaction and with EDFs
that are specifically fitted with primary consideration being the tensor
component. The calculations were done for five different orientations of the
deformed target, and with a full range of impact parameters for each orientations.
This allows for producing the fragment yield distributions, which in turn
tells us the dominant shell effects for the reaction.
As for the EDFs employed, we chose the SLy5 and SLy5t parametrizations,
which only differ by the addition of the tensor form in SLy5t without
refitting the other terms. In addition we have employed three TIJ EDFs that
were particularly constructed with tensor term in mind.

The charge and neutron distributions for the $^{48}$Ca+${}^{249}$Bk system, obtained using three different TIJ parametrizations, demonstrate a distinct difference in quasifission products between T31 and T44, T62 sets. Specifically, the T31 parametrization, characterized by a negative isovector coupling constant $C_1^J$ and a zero isoscalar coupling constant $C_0^J$, exhibits a pronounced shift in the neutron yield distribution, with the yield peak moving towards the neutron magic number $N = 126$ when compared to the T44 and T62 parametrizations. The yield peaks and distributions for the SLy5t and T31 parameter sets in the $^{48}$Ca+$^{249}$Bk collisions system closely resemble each other, reflecting their proximity in the $C_0^J$ -- $C_1^J$ parameter space. In contrast, the SLy5, T44, and T62 parameter sets, while differing significantly in $C_0^J$ and $C_1^J$, still produce distributions with considerable similarity. The comparison among SLy5, SLy5t and TIJ results suggests that the influence of tensor forces on the prominence of shell effects in quasifission is sensitive only in specific regions  of the $C_0^J$ -- $C_1^J$ parameter space while remaining insensitive in others.
The narrow region containing SLy5t and T31 within the parameter space, characterized by a negative isovector tensor $C_1^J$, may be crucial for a more prominent determination of the shell effect, warranting further investigation.
This sensitivity also motivates a more robust approach to parameter exploration consistent with the underlying statistical uncertainties on the effective interaction, as was studied recently for fusion reactions in TDHF~\cite{godbey2022}, though the computational cost will likely require the construction of surrogate models to sufficiently sample the space~\cite{bonilla2022}.

The $^{48}\mathrm{Ti}+{}^{238}\mathrm{U}$ collision system also exhibits a pronounced spherical shell effect when tensor forces are included, but it differs from the $^{48}$Ca+$^{249}$Bk system by producing narrower yield distributions, in addition to shifting peak positions. A comparison between the TIJ parametrizations of the $^{48}\mathrm{Ti}+{}^{238}\mathrm{U}$ collision system, particularly T31 and T62, exhibits enhanced shell effects through peak position shifts. Furthermore, the charge distribution using the SLy5t force, shows much better agreement with experimental results than that obtained with the SLy5 force.  Similarly, the distributions derived from the T31 interaction align more closely with experimental data compared to those from the T62 interaction.

In conclusion, the influence of tensor forces on shell effects, as observed through yield distributions, manifests in two primary ways: the shifting of peak positions and the alteration of distribution widths. The tensor force not only impacts shell evolution in nuclear structure but also plays a significant role in quasifission dynamics. While the distribution of quasifission products remains insensitive to variations in coupling constants across most regions, a specific region encompassing SLy5t and T31 exhibits more pronounced shell effects, resulting in significant shifts or narrowing of the distributions. This highlights the necessity for further investigation into the details of the EDFs and the role of tensor forces in nuclear collision processes.

\begin{acknowledgments}
This work has been supported by the Strategic Priority Research Program of Chinese Academy of
Sciences (Grant No. XDB34010000), the National Natural Science Foundation
of China (Grant Nos. 12375127 and 11975237), the U.S. Department of Energy under Grant Nos.
DE-SC0013847 (Vanderbilt University), DE-SC0013365, and DE-SC0023175 (Michigan State University). The computations in present work have been performed on the C3S2 computing center in Huzhou University and HPC cluster in Beijing PARATERA Tech Ltd.
\end{acknowledgments}


\bibliography{VU_bibtex_master.bib}

\end{document}